\documentclass[11pt,draftcls]{IEEEtran}{\onecolumn}
\usepackage{epsfig}
\usepackage{amsmath,epsfig,dsfont}
\usepackage{subfigure}
\usepackage{algorithm}
\usepackage{algorithmic}
\usepackage{amssymb,bm}
\usepackage{amsfonts}
\usepackage{stfloats}
\usepackage{multirow}
\usepackage{cite}
\ifCLASSINFOpdf
\usepackage{graphicx}
\else
\fi
\begin{document}
\title{Mobile Data Transactions in Device-to-Device Communication Networks: Pricing and Auction}

\author{Jingjing~Wang,~\IEEEmembership{Student Member,~IEEE,}
        Chunxiao~Jiang,~\IEEEmembership{Senior Member,~IEEE,}
        Zhi~Bie,\\
        Tony~Q.~S.~Quek,~\IEEEmembership{Senior Member,~IEEE,}
        and~Yong~Ren,~\IEEEmembership{Member,~IEEE}
\thanks{J. Wang, C. Jiang, Z. Bie and Y. Ren are with the Department
of Electronic  Engineering, Tsinghua University, Beijing 100084,
P. R. China. E-mail: chinaeephd@gmail.com, \{jchx, reny\}@tsinghua.edu.cn, bz12@mails.tsinghua.edu.cn.}
\thanks{T. Q. S. Quek is with the Singapore University of Technology and Design, 8 Somapah Road, Singapore 487372. Email: tonyquek@sutd.edu.sg.}
}
\maketitle

\begin{abstract}
Device-to-Device (D2D) communication is offering smart phone users a choice to share files with each other without communicating with the cellular network. In this paper, we discuss the behaviors of two characters in the D2D data transaction model from an economic point of view: the data buyers who wish to buy a certain quantity of data, and the data sellers who wish to sell data through D2D network. The optimal price setting and purchasing strategies are desirable, and we give the analysis based on game theory.
\end{abstract}


\IEEEpeerreviewmaketitle

\section{Introduction}
Device-to-Device (D2D) communication has been proposed as a revolutionary paradigm to enhance the capacity of cellular networks~\cite{1}.
According to D2D theory, a cell area under the control of one base station is divided into several clusters, and mobile users in the same cluster are close enough to establish direct connections. This clustering concept can remarkably improve the performance of cellular networks, and  several aspects of the benefits have been investigated~\cite{2}, including higher spectral efficiency, enhanced total throughput, higher energy efficiency and shorter delay.

Since multimedia files, like wireless videos, have been the main driver for the inexorable increase in data transmission, experts have proposed many methods to deal with the problem. In retrospect, researchers proposed methods like: decreasing the cell size to improve the spectral efficiency, using additional spectrum and improving the physical-layer link capacity, but all these traditional methods are not satisfactory and may even bring other problems. Later, it was observed that mobile devices have large storage space. Following this idea, a novel architecture based on D2D communication was thoroughly analyzed in~\cite{11}.

In this paper, we explore the mobile users' behaviors when they buy or sell videos within a D2D cluster in terms of an economic point of view. We model the file distribution as a competition mechanism against the data service of base station, and thus the autonomous D2D mechanism is preferred, where the data transaction is managed by D2D users without manipulation of base station. Furthermore, in order to analyze the optimal selling and purchasing strategies of the users, we use game theory to model different situations. Specifically, we model the one buyer multiple sellers case as a Stackelburg game, and model the one seller multiple buyers case as an auction game. The ultimate goal is to maximize the utility of both sides simultaneously.

%\hfill what is this?
%\hfill September 20, 2014

% \subsection{Subsection Heading Here}
% \subsubsection{Subsubsection Heading Here}

\section{System model}

In the formulation of the channel condition, we assume that in a certain cluster $C_{i}$, a data buyer $B_{i}$ purchases files from a data seller $S_{i}$ with total transmission power $G_{i}$. Moreover, the distance between $B_{i}$ and $S_{i}$ is represented by $d_{i}$, and $D_{ij}$ denotes the distance between two neighbor D2D clusters' centers. Also, we assume that the channel gain is $H_{i}$, as well as $\sigma^2$ represents the additive white Gaussian noise (AWGN) power. Considering $M$ neighbor D2D clusters in a cellular network, the cluster interference power of neighbor cluster $C_{j}$ is denoted as $G_{I,j}$ with the corresponding channel gain $H_{I,j}$, $j=1,\ldots,M$. For $C_{i}$, the Signal to Interference plus Noise Ratio (SINR) is given by:
%\begin{equation}\label{SNRi}
%{\textrm{SNR}}=\frac{GH}{\sqrt{d}\sigma^2}.
%\end{equation}
\begin{equation}\label{SNRi}
\textrm{SINR}=\frac{{{G}_{i}}H_{i}/\sqrt{{{d}_{i}}}}{{{\sigma }^{2}}+\sum\limits_{j=1}^{M}{({G_{I,j}}H_{I,j}/\sqrt{{{D}_{ij}}}})}.
\end{equation}
The total channel bandwidth available for data transactions can be deemed as $W$. Then, relying on Shannon formula, the maximal achievable bit rate between $B_{i}$ and $S_{i}$ follows:
%\begin{equation}\label{Ri}
%R=W\log_2\left(
%1+\frac{GH}{\sqrt{d}\sigma^2}
%\right).
%\end{equation}
\begin{equation}\label{Ri}
{{R}_{i}}=W{{\log }_{2}}\left(1+\frac{{{G}_{i}}H_{i}/\sqrt{{{d}_{i}}}}{{{\sigma }^{2}}+\sum\limits_{j=1}^{M}{({G_{I,j}}H_{I,j}/\sqrt{{{D}_{ij}}})}}\right).
\end{equation}

The total nodal delay is calculated as the sum of processing, queuing, transmission and propagation time. Relying on the delay model in~\cite{103}, which approximatively maps the network architecture into bi-directional graphs, the system delay can be denoted as:
\begin{equation}\label{D}
D(Y,V)= \beta(O) \frac{\sqrt{Y+V}}{\sqrt{\log{(Y+V)}}}\textrm{ ms},
\end{equation}
where the $Y$ represents the number of users (buyers and sellers) in a data transaction model, while $V$ denotes the number of other users in the same cellular network who are simultaneously transmitting data. Furthermore, $\beta(O)$ is a function of the size of transaction data $O$.

\section{Transaction Model}

In this section, we will discuss the data transaction problem in two different situations. The first situation consists of one data buyer purchasing from several data sellers, where the buyer can purchase different coding layers of a video from different sellers and combine these streams during the decoding process~\cite{200}. The second model consists of one seller selling data to multiple buyers. The major difference between the two models is whether the buyer or the seller takes the first move in the game.

\subsection{Initiating with the Buyer}

In this subsection, we consider the transaction model of one data buyer $B$ purchasing multimedia files from $N$ data sellers, namely, $\{S_1,\ldots,S_N\}$ in D2D cluster $C_{i}$. Then, the SINR and the bit rate between $B$ and $S_n$ can be calculated by Eq.~(\ref{SNRi}) and Eq.~(\ref{Ri}). Given the the total available bandwidth $W$, which will be evenly allocated to all users. The maximal achievable bit rate of the buyer is given by:
\begin{equation}
R_B=\frac{W}{N+1}\sum_{n=1}^N\log_2\left(1+\frac{{{G}_{n}}H_{n}/\sqrt{{{d}_{n}}}}{{{\sigma }^{2}}+\sum\limits_{j=1}^{M}{({G_{I,j}}H_{I,j}/\sqrt{{{D}_{ij}}})}}\right).
\end{equation}

Due to transmission delay, the strategies of the buyer and the sellers would be announced sequentially, which is beneficial of constructing a Stackelburg game model~\cite{simaan1976stackelberg}, where a data buyer takes the first step to send a detect signal, in order to inform other mobile users in the same cluster what particular video file he/she wants. Next, the potential sellers send back their prices along with the channel conditions. Then, the data buyer decides how much the transmission power he/she intends to buy from different sellers, respectively.

A data buyer can choose to buy from part of the sellers or all the sellers, and even he/she is capable of selecting service providers, if that way can lead to a higher reward.

\subsubsection{Caching Conditions}

Whether a seller has stored a copy of the file or not can affect his/her reward considerably, because this determines whether the seller needs to download the file from service provider before transmitting it to the buyer.

In~\cite{17}, Zipf distribution has been verified as a good model to measure the popularity of a set of video files. Thus, we assume that there are totally $K$ popular videos in this region within a month, and the popularity of the videos is subject to Zipf distribution with the parameter $\eta$. Then, the required frequency of the $i^{th}$ popular video file can be denoted as:
\begin{equation}
{{\lambda }_{i}}=\frac{1/{{i}^{\eta }}}{\sum\nolimits_{k=1}^{K}{1/{{k}^{\eta }}}}, 1\le i\le K.
\end{equation}
Meanwhile, we assume that the caching of the files are uniformly distributed among the users, which means that for every internal storage unit of a seller's device, the probability that it is used to store one of the $K$ files is $1/K$.

Given that a seller can store $\tau$ different popular video files of this month in his/her internal storage, there are $C_{K}^{\tau }$ possible different storage modes. Furthermore, we denote one of the possible storage mode of the seller as $\Gamma=\{\Gamma_1,\ldots,\Gamma_\tau\}$, where $\Gamma_i$, $i=1,\ldots,\tau$, represents the rank order among $K$ popular video files. Thus, the possibility of one of $C_{K}^{\tau }$ storage modes can be given by:
\begin{equation}
\textrm{Pr}(\Gamma )=\frac{\tau !(K-\tau )!}{K!},
\end{equation}
and the probability of this seller having a cached copy of the wanted file in terms of the $\Gamma$ storage mode follows:
\begin{equation}
\textrm{Pr}(\textrm{wanted}|\Gamma)= \sum_{i=1}^\tau \lambda_{\Gamma_i}.
\end{equation}
Relying on the total probability formula, the probability that a seller has a cached copy of the wanted file can be constructed as:
\begin{equation}\label{PrCache}
\textrm{Pr}(\textrm{Cache})=\sum_{\Gamma} \left[\textrm{Pr}(\textrm{wanted}|\Gamma) \cdot \textrm{Pr}(\Gamma ) \right].
\end{equation}

\subsubsection{Reward Functions}

The data buyer $B$ in cluster $C_{i}$ gains reward by successfully receiving the file with a good quality measured by the maximal achievable transmission rate. Specifically, $B$ has to pay the price $p_n (1 \leq n \leq N)$ for a unit of transmission power of $N$ sellers, respectively. Therefore, the data buyer's reward function can be formulated as:
\begin{equation}\label{PhiB-multiple}
\Phi_B\!=\!\frac{W}{N+1}\sum_{n=1}^N \log_2\left(1\!+\!\frac{{{G}_{n}}H_{n}/\sqrt{{{d}_{n}}}}{{{\sigma }^{2}}+\sum\limits_{j=1}^{M}{({G_{I,j}}H_{I,j}/\sqrt{{{D}_{ij}}})}}\right)\cdot \xi_R-D(Y,V)\cdot\xi_D
-\sum_{n=1}^N p_n\cdot G_n,
\end{equation}
where the coefficient $\xi_R$ represents the unit reward in terms of the maximal achievable bit rate, as well as $\xi_D$ denotes the unit loss measured by the system transmission delayed, which can be deemed as normalizing weight parameters for both $R_B$ and $D(Y,V)$. $G_n$ represents the traded transmission power that the seller $S_n$ transmits to $B$. The system delay $D(Y,V)$ is defined in Eq.~(\ref{D}).

In the following, we formulate the sellers' reward function. We introduce parameters $c_n$, $n=1,\ldots,N$, to represent the unit cost for relaying data between $B$ and $S_n$, which is determined by the characteristics of the sellers' devices. Thus, the utility function of $S_n$ can be defined as:
\begin{equation}\label{PhiS}
\begin{aligned}
\Phi_{S_n}=~&\textrm{Pr}(\textrm{Cache})\cdot(p_n-c_n)\cdot G_n+ \\
&\left[1-\textrm{Pr}(\textrm{Cache})\right]\cdot(p_n-c_n-s)\cdot G_n,
\end{aligned}
\end{equation}
where $s$ is the unit cost of downloading files from the nearest base station.
Additionally, we has a maximum cellular users constraint, i.e., $Y+V \leq \Omega$.
\subsubsection{Optimal Strategies}

In a Stackelburg game, the optimal transaction strategies for both sides exist and can be obtained using backward induction. The last move is for the data buyer to determine the optimal transmission power $G_n$ acquired from seller $S_n$ to maximize his/her reward $\Phi_B$. Moreover, we assume that the possible power of a received video is continuous~\cite{15}.

Hence, we can obtain $G_n$ by letting the first-order derivative of $\Phi_B$ with respect to $G_n$ be zero, which leads to
\begin{equation}\label{Qi}
{{G}_{n}}=\frac{W\xi_{R}}{(N+1){{p}_{n}}\ln 2}-\frac{\sqrt{{{d}_{n}}}\left[{{\sigma }^{2}}+\sum\limits_{j=1}^{M}\left({{G_{I,j}}H_{I,j}/\sqrt{{{D}_{ij}}}}\right)\right]}{H_{n}}.
\end{equation}

As the second last move, the sellers need to decide the optimal price $p_n$ in terms of the transmission power $G_n$ purchased by the buyer. We substitute $G_n$ in Eq.~(\ref{PhiS}) with Eq.~(\ref{Qi}), and let the first-order derivative with respect to $p_n$ equal zero, we have the optimal price of the data seller $S_{n}$:
%\begin{equation}
%\frac{W{{\xi }_{R}}[{{c}_{n}}+s(1-\Pr (\textrm{Cache}))]}{(N+1)p_{n}^{2}ln2}\!=\!-\!\frac{\sqrt{{{d}_{n}}}[{{\sigma }^{2}}\!+\!\sum\limits_{j=1}^{M}({{G_{I,j}}H_{I,j}/\sqrt{{{D}_{ij}}}})]}{H_{n}}.
%\end{equation}
%Therefore, the optimal price of the data seller $S_{n}$ should be set as:
\begin{equation}\label{pi}
p_n=\sqrt{\frac{WH_{n}\xi_R[c_n+s(1-\Pr(\textrm{Cache}))]}{(N+1)\sqrt{d_n}\left[\sigma^2+ \sum\limits_{j=1}^{M}\left({{G_{I,j}}H_{I,j}/\sqrt{{{D}_{ij}}}}\right)\right]\ln{2}}}.
\end{equation}
Substituting $p_n$ in Eq.~(\ref{Qi}) with Eq.~(\ref{pi}), we can achieve the optimal transmission power $G_n$.

\subsection{Initiating with the Seller}

In this situation, if a seller wants to sell data to nearby users, he/she takes the first step to send a probe signal to announce the availability of stored data resources. The seller would prefer to maximize his/her reward by transmitting popular video files in terms of a fierce competition among the potential multiple data buyers.

We model this decentralized process as alternative ascending clock auction (ACA-A) game, because it guarantees an efficient and cheat-proof allocation~\cite{chen2010spectrum}. In our model, there are $N$ buyers, namely $B_n$, $n=1,\ldots,N$, competing for high quality data from one seller $S$.

Similarly, the SINR and maximal achievable bit rate are given by Eq.~(\ref{SNRi}) and Eq.~(\ref{Ri}). Then, the reward function of the buyers at clock step $t$ can be deemed as:
\begin{equation}
\Phi_{B_n}^t(p^0,G_n)=R_{B_n}^{t}\cdot \xi_R-D(Y,V)^{t}\cdot\xi_D-p^{t}\cdot G_{n}^{t},
\end{equation}
where, $p^{t}$ represents the price of unit transmission power charged by $S$.
Suppose that the seller wants to sell total power of $G$. In the beginning of the auction, $S$ sets up the clock index $t=0$, regulates time-step size $\delta >0$, and announces the initial price $p^0$ to all the D2D users in the cluster $C_{i}$. Then, each potential buyer $B_n$ offers his/her optimal bid $G_n^0$ by computing
\begin{equation}
G_n^0=\arg \max_{G_n} \Phi_{B_n}\left(p^0,G_n\right).
\end{equation}
Then, the seller sums up all the bids $G_{total}^0=\sum_n {G_n^0}$, and compares $G_{total}^0$ with $G$. If $G_{total}^0 \leq G$, the seller concludes the auction. Otherwise, the seller sets $p^{t+1}=p^t+\delta$, $t=t+1$ and announces the price $p^t$ to all the buyers who bid for the next time. Thus, each buyer offers his/her optimal bid $G_n^{t+1}$ by calculating
\begin{equation}
G_n^{t+1}=\arg \max_{G_n} \Phi_{B_n}\left(p^{t+1},G_n\right).
\end{equation}

After collecting all the bids, the seller computes the total bid $G_{total}^{t+1}$. The seller continues the auction until $G_{total}^{t+1} \leq G$. At every clock $t$, the seller computes the cumulative clinch $\chi_n^t$, which is the amount of data that the buyer $B_{n}$ is guaranteed to win at clock $t$:
\begin{equation}
\chi_n^t=\max \left(0,G-\sum_{j \neq n}{G_j^t}\right),
\end{equation}
and the buyer will purchase this $\chi_n^t$ quantity of power with price $p^t$. This has been proved to be a cheat-proof scheme.

Let the final clock index be $T$. As the price $p$ increases discretely, we may have $G_{total}^{T} < G$, in which case the power resource is not fully utilized. We adopt a proportional rationing scheme in order to make sure that $G_{total}^{T} = G$ ~\cite{ausubel2004efficient}. Therefore, the final allocated transmission power can be denoted by:
\begin{equation}\label{Qi_final}
G_{n,final}=G_n^{T}+\frac{G_n^{T-1}-G_n^{T}}{\sum\limits_{i=1}^{N}{G_i^{T-1}}-\sum\limits_{i=1}^{N}{G_i^{T}}} \left(G-\sum\limits_{i=1}^{N}{G_i^{T}}\right),
\end{equation}
which satisfy $\sum_n{G_{n,final}}=G$.

We set $\chi_n^{T}=G_{n,final}$, and the total payment of buyer $B_n$ is given by:
\begin{equation}
P_n=\chi_n^0p^0+\sum_{t=1}^{T} p^t\left(\chi_n^t-\chi_n^{t-1}\right).
\end{equation}

In this model, the reward of the seller can be formulated as:
\begin{equation}
\Phi_{S}=\sum_{n=1}^N{\left(P_n-c_n G_{n,final}\right)},
\end{equation}
where $c_n$ accounts for the unit transmission cost.

\section{Simulation Results}

In this section, we will show the variation tendency of the sellers' and buyers' reward functions and their optimal choices in different models.

\subsection{Initiating with the Buyer}

In our simulation, we set the data buyer at the origin (0,0) and let other mobile users uniformly distribute in a cell. The radius of a cell is $500$m, as well as the radius of a D2D cluster is set as $100$m. Potential sellers, i.e., $N=10$, are uniformly distributed in a D2D cluster. Besides, the number of clusters, which are randomly scattered in a cell, is $M=10$. We set the noise power spectrum density to be $-174$dBm, and the total available bandwidth for D2D data transaction is $5$MHz under an ideal channel gain $H=1$. Moreover, for all users, the maximal transmit power is $100$mW, i.e., $G_{n}\leq 100$mW, $G_{I,j}\leq 100$mW, and the normalized weighting parameters $\xi_R=3\times10^{-5}$/bps as well as $\xi_D=10^{-1}$/ms in terms of $\beta(O)=20$ under the assumption of the same packet size $O=1$MB. The data seller's unit transmission power cost $c$ is uniformly distributed in $[0.1, 0.5]$/mW. Finally, we set the probability that the seller has stored a copy of the wanted file to be $\textrm{Pr}(\textrm{Cache})=0.3, 0.4, 0.5$, respectively.

\begin{figure}
\begin{center}
\includegraphics[width=0.6\textwidth]{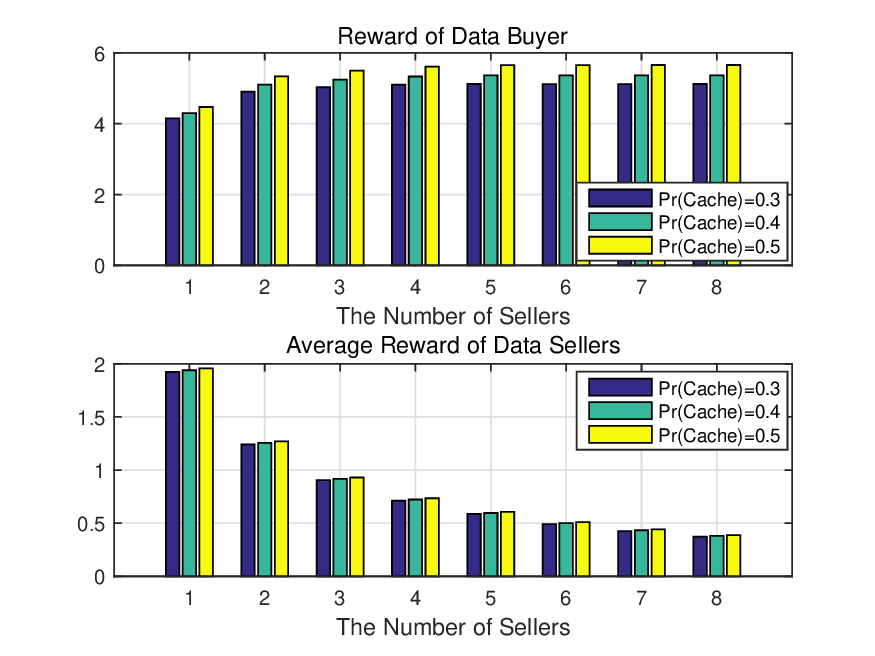}
\end{center}
\caption{Variation tendency of the reward functions for the buyer and sellers versus the increasing sellers' number.}\label{MultipleU}
\end{figure}

\begin{figure}
\begin{center}
\includegraphics[width=0.6\textwidth]{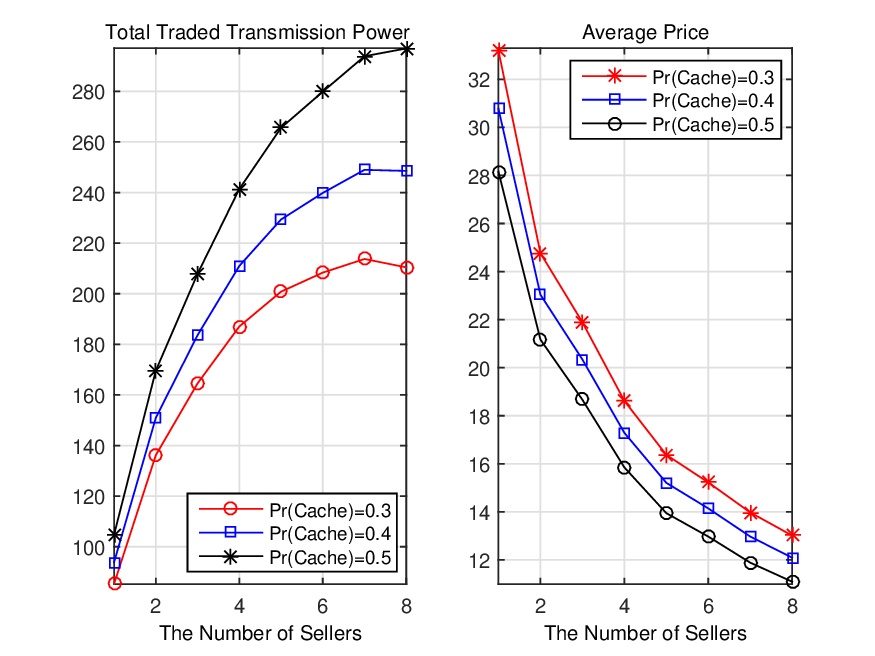}
\end{center}
\caption{Variation tendency of the optimal choices of the buyer and sellers versus the increasing sellers' number.}\label{MultipleE}
\end{figure}

\begin{figure}
\begin{center}
\includegraphics[width=0.6\textwidth]{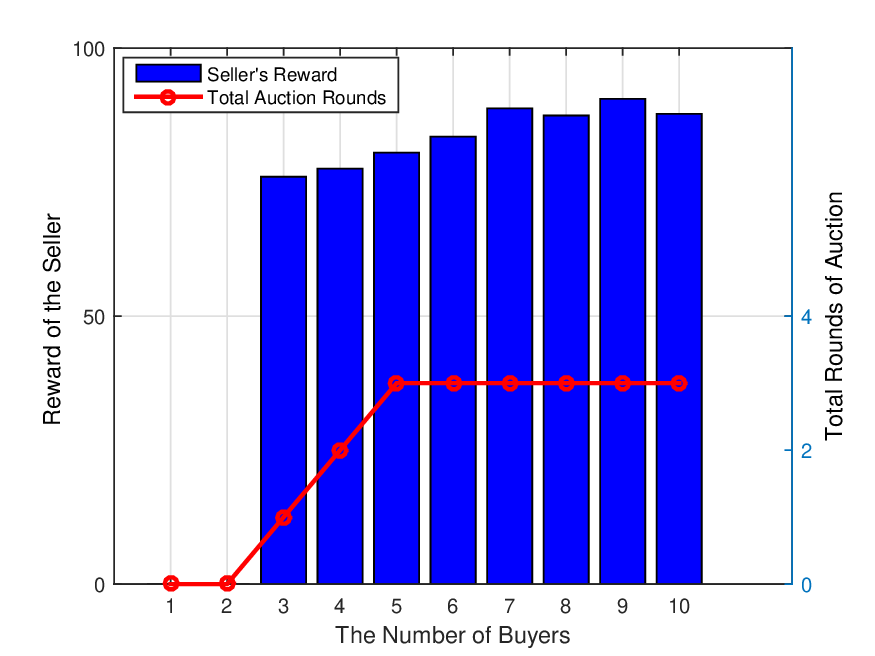}
\end{center}
\caption{Variation tendency of the seller's reward function and the total rounds of the auction with respect to the number of buyers.}\label{auction1}
\end{figure}
From Fig.~\ref{MultipleU} and Fig.~\ref{MultipleE}, we can conclude that as the number of potential sellers increases, the average price of sellers decreases, while the total traded transmission power improves. Furthermore, the data buyer's reward increases at the beginning, and then keeps constant. That is because when there are more sellers, they may compete with each other and lower the price. However, when the buyer purchase from superfluous buyers, system delay start to play a critical role in reducing the rewards. Also, with the improving of the probability of having a wanted cached video copy, the sellers tend to lower their prices, which contributes to higher rewards.
\subsection{Initiating with the Seller}

The seller's price is set to be $5$ at the first clock step, as well as the total available transmission power of the seller is $G=100$mW. Other simulation parameters are the same as above. The variation tendency of the seller's reward is shown in Fig.~\ref{auction1} in terms of different number of buyers.

With less than $3$ buyers, the transaction is finished in the first round and the seller choose to utilize the data resource him/herself. However, given more buyers, the seller allocates the power resources relying on the ACA-A auction scheme. As the number of buyers increases, the auction takes more rounds, as well as the traded price tends to be higher, which leads to higher rewards of the seller.

\section{Conclusion}

In this paper, we established two data transaction models in D2D communication networks. A Stackelburg game as well as an ACA-A auction model were proposed for one-buyer/multiple-seller situation and one-seller/multiple-buyers situation, respectively. Moreover, theocratical analysis and numerical simulations were conducted in order both to achieve optimal trading strategies and to provide a new research method for D2D users' behavior.

%\section*{Acknowledgment}

%IEEEhowto:kopka

\bibliographystyle{IEEEtran}
\bibliography{IEEEabrv,ref}
%\begin{thebibliography}{1}
%\bibitem{}
%\end{thebibliography}

\end{document}